**Impact of the energetic landscape on polariton condensates' propagation along a coupler**

Elena Rozas, Johannes Beierlein, Alexey Yulin, Martin Klaas, Holger Suchomel, Oleg Egorov, Ivan A. Shelykh, Ulf Peschel, Christian Schneider, Sebastian Klembt, Sven Höfling, M. Dolores Martín* and Luis Viña

Dr. E. Rozas
Dept. Física de Materiales, Universidad Autónoma de Madrid, 28049 Madrid, Spain
Instituto Nicolás Cabrera, Universidad Autónoma de Madrid, 28049 Madrid, Spain

J. Beierlein,
Technische Physik, Wilhelm-Conrad-Röntgen-Research Center for Complex Material Systems, and
Würzburg-Dresden Cluster of Excellence ct.qmat,
Universität Würzburg, Am Hubland, D-97074 Würzburg, Germany

Dr. A. Yulin,
National Research University of Information Technologies, Mechanics and Optics (ITMO University), Saint-Petersburg 197101, Russia

Dr. M. Klaas
Technische Physik, Wilhelm-Conrad-Röntgen-Research Center for Complex Material Systems, and
Würzburg-Dresden Cluster of Excellence ct.qmat,
Universität Würzburg, Am Hubland, D-97074 Würzburg, Germany

H. Suchomel
Technische Physik, Wilhelm-Conrad-Röntgen-Research Center for Complex Material Systems, and
Würzburg-Dresden Cluster of Excellence ct.qmat,
Universität Würzburg, Am Hubland, D-97074 Würzburg, Germany

Dr. O. Egorov
Institute of Condensed Matter Theory and Solid State Optics, Abbe Center of Photonics,
Friedrich-Schiller-Universität Jena, D-07743 Jena, Germany

Prof. I. A. Shelykh,
Faculty of Physics and Engineering, ITMO University, 197101 St. Petersburg, Russia
Science Institute, University of Iceland, IS-107 Reykjavik, Iceland,

Prof. U. Peschel
Institute of Condensed Matter Theory and Solid State Optics, Abbe Center of Photonics,
Friedrich-Schiller-Universität Jena, D-07743 Jena, Germany

Dr. C. Schneider
Technische Physik, Wilhelm-Conrad-Röntgen-Research Center for Complex Material Systems, and
Würzburg-Dresden Cluster of Excellence ct.qmat,
Universität Würzburg, Am Hubland, D-97074 Würzburg, Germany

Dr. S. Klembt
Technische Physik, Wilhelm-Conrad-Röntgen-Research Center for Complex Material Systems, and
Würzburg-Dresden Cluster of Excellence ct.qmat,
Universität Würzburg, Am Hubland, D-97074 Würzburg, Germany

Prof. S. Höfling
Technische Physik, Wilhelm-Conrad-Röntgen-Research Center for Complex Material Systems, and
Würzburg-Dresden Cluster of Excellence ct.qmat,
Universität Würzburg, Am Hubland, D-97074 Würzburg, Germany
SUPA, School of Physics and Astronomy, University of St Andrews, St Andrews KY16 9SS, United Kingdom

Prof. M. D. Martín
Dept. Física de Materiales, Universidad Autónoma de Madrid, 28049 Madrid, Spain
Instituto Nicolás Cabrera, Universidad Autónoma de Madrid, 28049 Madrid, Spain
E-mail: dolores.martin@uam.es

Prof. L. Viña
Dept. Física de Materiales, Universidad Autónoma de Madrid, 28049 Madrid, Spain
Instituto Nicolás Cabrera, Universidad Autónoma de Madrid, 28049 Madrid, Spain
Instituto de Física de la Materia Condensada, Universidad Autónoma de Madrid, 28049 Madrid, Spain

Polariton condensates' propagation is strongly dependent on the particular energy landscape the particles are moving upon, in which the geometry of the pathway laid for their movement plays a crucial role. Bends in the circuit's trajectories affect the condensates' speed and oblique geometries introduce an additional discretization of the polaritons' momenta due to the mixing of short and long axis wavevectors on the propagating eigenvalues. In this work, we study the nature of the propagation of condensates along the arms of a polariton coupler, by a combination of time-resolved micro-tomography measurements and a theoretical model based on a mean field approximation where condensed polaritons are described by an equation for the slow varying amplitude of the polariton field coupled to an equation for the density of incoherent excitons.

# 1. Introduction

In the last couple of decades, the physics of exciton – polaritons (in the following, just polaritons) in semiconductor microcavities has been a very active research field, investigating both fundamental physical properties of quantum fluids of light and potential applications envisioning new optical devices. The ultimate properties of polaritons are based on the strong coupling of radiation and matter achieved in these microcavities. Since their first experimental demonstration in 1992, [1] a titanic effort has been made to exploit their remarkable light – matter qualities with the purpose of finding new mechanisms for the control and manipulation of light. Using these particles, it has been possible to observe Bose – Einstein condensation, [2] superfluid motion, [3,4] and the persistence of vortices. [5] The development of different approaches to engineer couplings between spatially separated condensates [6] has made possible the experimental demonstration of a polariton simulator, exploiting bosonic stimulation to deterministically find the ground state of the XY spin glass for arbitrary lattices. [7] Moreover, during the last decade, a unique control over the realization and manipulation of polariton lattices has been attained and several experimental demonstrations of the so-called photonic graphene have been published. In the early ones, only the linearity of the polariton dispersion around K and K' points of the reciprocal space was proven under optical [8] and electrical injection, [9] while in the later ones, the evidence of topological states in this kind of lattices have been demonstrated. [10]

Concurrently, diverse polariton devices have been proposed, such as polariton lasers, [11-13] optical gates, [14] transistors, [15-17] spin-based elements [18-21] and integrated circuits. [22 - 24]. Yet, a close look at the dynamics of propagating polariton condensates along **non-straight** circuits is lacking, a void we start filling with this work. We find that the potential landscape that polaritons experience depends on their position along the circuit and that this fact is not linked with a wedge gradient, habitual in semiconductor microcavities. The change in the orientation of the longitudinal direction for propagation at different positions along the circuit has a remarkable impact on the polariton momentum. Additionally, we find a deceleration when polaritons turn at the bends of the circuits where they propagate. Our experimental findings are supported by theoretical simulations based on the open and dissipative Gross-Pitaevskii equation describing the slow varying amplitude of the polariton field and a rate equation written for the density of the incoherent excitons.

## 2. Sample and experimental details

The microcavity waveguides have been grown by molecular beam epitaxy and processed by reactive ion etching. They consist of a λ/2 cavity sandwiched between two distributed Bragg reflectors of 23/27 pairs of alternating layers of $Al_{0.2}Ga_{0.8}As$/AlAs on the top/bottom. Three sets of 4 GaAs quantum wells (QWs), 7 nm wide, have been placed at the antinodes of the electromagnetic field confined inside the cavity. A Q-factor of ~ 5000 and a Rabi splitting of 13.9 meV have been determined experimentally by low power measurements. Couplers consisting of adjacent waveguides of different width and length have been repeatedly patterned throughout the surface and realized by etching down to the QWs. A sketch of one of these devices is shown in **Figure 1**. The input and output terminals, 42 μm long, are located at each edge of the doubly bent waveguide and rotated ±45º from the longitudinal direction. The structure studied here is 6 μm wide (w), the longitudinal part (L) is 100 μm long and the separation between the mirrored waveguides (d) is 1.5 μm. This gap between the coupler's arms is large enough to avoid any coupling between them and allows us to study the polariton dynamics in a single arm. The sample is optically pumped non-resonantly at 1.664 eV by means of a pulsed Ti:Sapphire laser (2 ps pulse duration, 82 MHz repetition rate) and under normal incidence. The excitation beam is focused down to 3 μm in one of the waveguides using a 20× microscope objective (NA = 0.40, f = 4 mm) that is also used to collect the photoluminescence (PL) emitted by the 1D waveguide, kept at 12 K inside a cold finger cryostat. For an excitation power above a certain threshold (12 kWcm$^{-2}$) polariton condensates propagate away from the excitation area.[25] Our experimental setup permits the consecutive measurement of the PL in both real and reciprocal (momentum) space, allowing us to follow the movement of polaritons in real space while simultaneously registering their speed (wavevector) and their energy dispersion. [26] To measure the polariton dispersion relations the PL has been spatially filtered, so that only the emission originating from the excitation area is collected. In this manuscript, we will concentrate on the study of the propagation of polariton condensates along one arm of the coupler only, leaving for a future publication the analysis of the coupling between the arms.

## 3. Local photonic landscape and propagation of polariton condensates

A characterization of the photonic landscape experienced by polariton condensates in their movement is necessary to fully understand polariton propagation and to improve the design of polaritonic circuits. For this purpose, the structure has been pumped at different locations along both arms of the coupler. Since the results obtained for both are essentially equivalent, we will concentrate on the results acquired in just one of them and on three selected positions that summarize the results obtained along the whole structure. Position 1 and 3 are placed in the input/output terminals, respectively, while position 2 is located in the central region. We measure the PL spectrum both along the x- and y- direction (axes defined in Figure 1). Note that for position 2, these axes are oriented along the longitudinal and transversal directions of the waveguide, respectively, while the terminals are rotated ±45º with respect to these axes. The dispersion relations at the three locations are summarized in **Figure 2**. Only the PL from an area of 60 μm$^2$ around the excitation spot has been collected as the emission arising from the rest of the device has been filtered out using a spatial filter. Panels (a), (c) and (e) show the PL angular distribution along $k_x$ (measured at $k_y = 0$) for a low excitation power, 0.6 kWcm$^{-2}$, at positions 1-3.

Let us start by analyzing the emission at position 2. In the case shown in panel (c), the dispersion is produced entirely by the contribution of polaritons distributed along the longitudinal direction of the waveguide, which coincides with the direction of detection. Polaritons can freely expand along the longitudinal direction of the guide so that no restriction is observed in the angular detection. When polaritons annihilate, photons are emitted at all possible angles along the x axis. Thus, the entire lower polariton branch (LPB) and the rest of sub-bands are observed, exhibiting continuous energy bands in the full range of $k_x$. By contrast, the discontinuity in the refractive index in the transverse direction, created by the difference between the cavity and air, yields a complete lateral confinement and hence to a full quantization of the transverse momentum. The allowed values of this wavevector are approximately given by k = j π/w, where w is the waveguide's width and j is the quantization number, integer and positive, that identifies each sub-band in the dispersion. Panel (d) demonstrates a completely discretized dispersion along $k_y$, due to the high confinement produced by the narrowness of the waveguide. The emission is split into several sub-bands whose antinodes are clearly visible. The superposition of the dispersions in panels (c) and (d) is responsible for the dispersion measured in the terminals of the waveguides, as it will be shown below. It is worthwhile to mention that the dispersions shown in panels (c) and (d) are symmetric with respect to an inversion from k to - k, as expected from the symmetry of the waveguides and the excitation conditions at the center of the coupling region. A full tomography ($k_x$ vs $k_y$ as a function of the energy) for this position can be found in the supporting information (see **Video S1**).

Now, we analyze the dispersions along $k_x$ at the terminals. Conversely to the symmetry discussed for the dispersions at the coupling region, the multiband structure between 1.584 and 1.598 eV, observed in the emission at the input terminal, position 1 [panel (a)], exhibits a peculiar asymmetry between negative and positive values of $k_x$, with a larger population observed in the former one. This dispersion relation, since the waveguide is rotated 45º with respect to the detection axes, can be understood as a linear combination of those associated to the longitudinal and transversal eigenvectors of the waveguide. [27] Furthermore, the position of the laser beam near the waveguide edge plays an important role as it introduces a strong asymmetry in the dispersion. An additional wavevector discretization along the longitudinal direction emerges due to the finite distance to the edge of the waveguide, resulting in an accentuated energy discontinuity at negative values of the momentum. The quasi-confinement of polaritons between the edge of the waveguide and the exciton reservoir potential barrier at the excitation spot is responsible for the larger intensity obtained at negative values of $k_x$ in the dispersion. For positive values there is no obstacle to the polariton expansion since they can travel at ease along the waveguide, so a more continuous band for $k_x > 0$ is found. Additional experimental evidence (not shown here) reveal that the closer to the waveguide's edge the excitation, the more pronounced the asymmetry of the emission. This is due to the fact that the allowed values of the discrete longitudinal wavevector are further separated because of the increasing proximity of the edge and also to the polariton population getting trapped in a smaller area of the waveguide (smaller the closer to the edge of the structure). The excitation at position 3 [panel (e)] obtains a similar dispersion relation to that shown in panel (a). However, in this case, the waveguide is rotated – 45º, and therefore the emission shows discrete energy states at positive values of $k_x$.

The corresponding dispersions relations along $k_y$ (at $k_x = 0$) for positions 1 and 3 are shown in panels (b) and (f), respectively. Panel (b) exhibits a similar distribution to that of panel (a) due to the additional confinement created by the edge of the structure, which in this case also results in a discretization for negative values of $k_y$. Due to the opposite angle of rotation, with respect to x, of the output terminal to that of the input one, panel (f) shows the reverse behavior to that of panel (b) with discrete values predominantly for $k_y > 0$. The tomography measured at one of the terminals can be found in the supporting information (see **Video S2**).

The superposition of both longitudinal and transverse wavevectors on $k_x$ and $k_y$, together with the additional confinement at the end of the input and output terminals, provides a full understanding of the polariton dispersion relations, explaining also the smoother discretization of the energies at the terminals as compared to that observed in panel (d). To fully clarify the propagation of polaritons and the influence of mode mixing, a comparison between complete experimental tomographic maps of the dispersion relations and the results of the simulations is presented in the supporting information (see **Figure S1**).

To fully understand the experimental results, we performed numerical simulations of the polariton dynamics. For this we adopted a model describing the polaritons by a set of coupled equations for the slow varying amplitudes of polaritons and incoherent excitons. The excitons affect the polaritons in two ways: they create a repelling potential and also introduce an effective gain favoring the creation of polaritons at the bottom of the LPB. Physically, this latter process can be seen as a phonon-assisted relaxation along the dispersion that also leads to the depletion of the exciton bath. The coupled equations describing the polaritons and the excitons read

$$i\hbar \partial_t \psi = -\frac{\hbar^2}{2m_{eff}} \nabla^2 \psi + i\frac{\hbar}{2}(Rn - \gamma)\psi + V(\vec{r})\psi + gn\psi + g_c|\psi|^2\psi + P_p(t,\vec{r})$$
$$\partial_t n = -(\Gamma + R|\psi|^2)n + P_n(t,\vec{r}). \qquad (1)$$

Here $\psi$ is the complex amplitude of the polaritons, $n$ the density of incoherent polaritons, $m_{eff}$ the effective mass of the polaritons, $\gamma$ their decay rate, $R$ the coefficient defining the rate of creation of polaritons from excitons, $V(\vec{r})$ the effective confinement potential created by the microstructuring of the sample, $g$ the coefficient accounting for the effective repelling potential created by the incoherent excitons, $g_c$ the nonlinear mutual interaction of the polaritons and $\Gamma$ the decay rate of the exciton bath. In our simulations, we assumed $\gamma$ to be a function of the coordinates such that within the polariton waveguide the losses are lower compared to those outside the waveguide, where the etching deteriorated the quality of the mirrors.

We should mention that **Equation 1** also describes the process of polariton condensation that we will consider later, and the creation of polaritons directly by the driving force $P_p(t,\vec{r})$, which accounts not only for the excitation by the external light but also the effect of fluctuations. In our experiments, the spectral width of the pulses is much smaller than the detuning of the pulses' frequency from the exciton resonance. Therefore, below the condensation threshold, the polaritons in the waveguides appear because of the fluctuations. To describe this effect, we followed an approach[28,29] based on a Langevin-kind equation. So in the equation for the polaritons, $P_p(t,\vec{r}) = i\hbar\sqrt{\gamma + Rn}\partial_t \psi_{st}$ is a stochastic term and $\psi_{st}$ is a Gaussian random function $\delta$-correlated in space and time. Let us emphasize that the intensity of the fluctuations is higher in the area irradiated by the incident beam where the concentration of the incoherent excitons, $n$, is higher. In this way the incident beam creates a localized source of incoherent polaritons.

The incident pulse also creates the exciton bath: its action on the excitons is accounted by the term $P_n(t,\vec{r})$ For our numerical simulations we took the pump in the form

$$P_n(t,\vec{r}) = a_n \exp\left(|\vec{r} - \vec{r}_p|^4/w_p^4\right) \exp\left(-\frac{2t^2}{T^2}\right) \qquad (2)$$

where the coefficient $a_n$ is the amplitude of the pulse, $w_p$ the radius of the excitation spot and $T$ the duration of the pulse.

For the simulations we took the following parameters $R = 0.01$ μm²/ps, $g_c = 6 \cdot 10^{-3}$ meV/μm², $g = 2g_c$, $\Gamma = 3.5\gamma$. The polariton mass $m_{eff} = 4.8 \cdot 10^{-5} m_e$ ($m_e$ is the electron mass) and the polariton dissipation rate $\gamma = 0.24$ ps$^{-1}$ were taken to fit the experimentally measured dispersion of polaritons and the quality factor of the polariton resonator.

The polaritonic waveguide is modelled by a coordinate dependent potential replicating its shape properly. The depth of the potential was chosen to be 10 meV, which is enough to provide a good confinement of the polariton field at all frequencies of interest and to reproduce the experimental conditions. In order to simplify the calculations and to avoid the back reflection of the waves at the boundaries of the simulation window we have assumed the losses outside the waveguide to be 15 times larger than inside. The exact values of this dissipation do not affect the propagation of the guided modes much because of the tight confinement of the modes in the waveguides.

We start now considering the linear case, when the pump is weak enough so that the effect of the exciton bath on the dynamics of the polaritons is negligible regarding the amplification and the blue shift of the polaritons. The density of the polaritons produced by the fluctuations is so low that the nonlinear interaction between polaritons is negligible, thus the results are scalable with the pump intensity. We have simulated the dynamics of the polaritons and calculated the spatio-temporal spectra of their field within a light collection area of the same size as that used in the experiments. Since our system is excited by a random force, we did a number of simulations and then averaged the spectra over 100 shots. These simulated dispersion relations are presented in **Figure 3** (a,b) when the coupler is excited at position 2. The dispersion relations are symmetric and comply very well with what was observed in the experiments, see Figure 2 (c,d).

As it has been discussed above, when the waveguides are excited at the output terminals, sufficiently close to their edges, the dispersion plots change considerably [typical experimental examples are presented in Figures 2 (a,b) and (e,f)]. Comparing these two sets of figures it is clearly seen that the symmetry $k_x = -k_x$, $k_y = -k_y$ is broken (the dispersions measured at positions 1 and 3 are not exactly the mirror images of each other since the parameters of the waveguides slightly vary in space). These effects are well reproduced by the numerical simulations, see Figure 3 (c,d). [30] To fully understand the origin of the asymmetry, further numerical simulations (labelled as 3[+]) were performed for an excitation spot situated closer to the end of the waveguide, which borne out that the symmetry breaking is due to the reflection of the polaritons at the edge of the waveguides. The results of these simulations, shown in Figure 3 (e,f), clearly show that the positions of the maxima get shifted slightly but, more importantly, the intensity asymmetry becomes much more pronounced. [30] Thus, we can conclude that there is an excellent agreement between the experiments and our theoretical predictions.

Increasing the excitation power up to 24 kWcm[-2], well above the condensation threshold, polariton condensates injected in the waveguide propagate all over the device. **Figure 4** presents the corresponding dispersion relations obtained for the same three excitation points discussed earlier. The dashed lines mark the bottom of the LPB band when a low pump power is employed. The emission energy is blueshifted (~5 meV) due to polariton-polariton interactions, which is a well-known signature of polariton condensation.[2] Above threshold, two drops of condensates are created, propagating away from the excitation spot towards opposite longitudinal directions.

We discuss again, for the sake of clarity, the case of excitation at the center of the coupling region, position 2. In this case, along the longitudinal direction of the waveguide, apart from some polaritons that stay at rest ($k_x \approx 0$), condensates propagate with $|k_x| \approx 2$ μm$^{-1}$ at ~ 1.588 eV [Figure 4 (c)]. In the transverse direction, condensates propagate mainly at an excited state with slightly higher energy, 1.589 eV, and lower momentum, $|k_y| \approx 1.5$ μm$^{-1}$, [Figure 4 (d)].

A more complicated situation arises at the input and output terminals. Additional to the frequent presence of polaritons halted at the excitation area by the potential created by the excitation beam, massive populations at positive and negative values of the momentum are observed in the interval $2.0 < |k_x| < 3.5$ μm$^{-1}$. At the input terminal, condensates travel from position 1 towards the center of the structure (position 2) exhibiting positive momenta while most of the condensed population travels towards the edge of the structure, and it is trapped, with negative values of the momenta, as demonstrated by the emission depicted in Figure 4 (a). The situation is inverted at the output terminal [panel (e)] due to the presence of the edge of the structure affecting now those polaritons travelling with $k_x > 0$. In contrast to the findings at the center of the coupler [panel (c)], where condensates travel mostly with a well-defined single mode, peculiar multi-mode dispersions characterize the polariton propagation at the terminals. A similar comment pertains when analyzing the behavior along $k_y$: panels (b) and (f), corresponding to the input and output terminals, respectively, exhibit the presence of up to 4 modes similarly populated at different energies, while panel (d) shows that at the coupling region polaritons propagate mostly at a single mode.

To get a deeper insight into the experimental results, a series of numerical simulations have been performed for the case when the pump exceeds the threshold value. In this case, the dynamics of the polaritons is governed by two processes: first the polaritons are excited by the fluctuations and then they are amplified due to the condensation of the polaritons. The spectra obtained in this case by the numerical simulations are shown in **Figure 5**, where we have increased the intensity of the pump so that the density of the generated incoherent excitons shifts the frequency of the polaritons to approximately 1.588 eV.

Let us start discussing the excitation at the center of the waveguide, Figure 5 (a,b). These new dispersions can be interpreted as follows. At the initial stage, after excitation, polaritons are created within the excitation spot; the interactions with the incoherent excitons blue shift the polaritons' dispersion relations at this spot up to 1.588 eV. These polaritons get further amplified due to the gain produced by the incoherent excitons and propagate away, left and right, from the excitation spot due to the exciton reservoir potential barrier, eventually leaving the area where the gain exists. Polaritons with the lowest energies have lower group velocity and stay in the gain area longer than those with higher energies, therefore experiencing the strongest amplification. This explains why the polariton's frequency range above the condensation threshold is narrower than that of polaritons created below threshold. It should also be mentioned that in the experimental dispersions there are relatively pronounced spectral patterns around 1.588 eV and small wavevectors [see Figure 4 (c,d)], whereas in the calculations these patterns are hardly visible [see Figure 5 (a,b)]. This discrepancy can be traced to the fact that in the experiments these spectral patterns are produced at short times after excitation and their brightness depend strongly on both the size of the emission collection area and the relaxation times of the polaritons and incoherent excitons. In the numerical simulations, it was checked that reducing the radius of the light collecting spot the brightness of these spectral features can be significantly increased.

The simulated dispersions when the waveguide is excited in the output terminal (position 3) are shown in Figure 5 (c,d). [30] It is clearly seen that the reflection from the waveguide edge breaks the symmetry in a similar manner as how it happens for an excitation power below the condensation threshold, previously discussed.

To conclude this part of the work, we can say that the influence of the symmetry breaking on the propagation of polaritons in semiconductor microcavities is demonstrated experimentally and that the results can be well explained within the standard mean field approximation applied to polariton dynamics.

**4. Influence of the bend on polariton propagation**

Polariton condensation in 1D-waveguides has been frequently reported under non-resonant excitation, showing the formation of condensates propagating out of the excitation area. [20, 25, 31] However, a locally induced variation in the trajectory of these condensates has important effects for their implementation in polariton circuits. To our knowledge, the condensate's response under these conditions has been scarcely studied in the literature, [32, 33] with none of the works paying attention to the momenta distribution of a propagating condensate. Here we address this by investigating the condensates' trajectory after excitation of the device in position 1 and analyzing the effect of the bend between the terminal and the central region on the condensate's wavevector. **Figure 6** (a) shows the time-integrated emission of condensates propagating along the device. One of the condensates, ejected to the left, gets trapped between the edge of the waveguide and the excitation spot, bouncing repeatedly between these positions. The condensate going to the right shows a remarkable decay of the emission intensity when arriving to the bend of the structure. However, as shown in the figure, beyond x ~ 15 μm the propagation is visible up to 80 μm (note that the intensity has been multiplied by a factor 3). The emission displays a peculiar propagation along the device: condensates travel with a combined mode pattern along the input terminal, shaped in two parallel paths, mostly at the edges of the waveguide, while in the central region they merge and exhibit a zig-zag path. These changes, both in intensity and behavior, evidence the impact of a bend on the propagation of the condensates. A detailed view of this peculiar propagation can be observed in the time-resolved emission shown in **Video S3** of the supporting information, where polaritons are injected in the central region and propagate towards the output terminal of a coupler.

To quantify in more detail the impact of a 45º rotation of the trajectory, we measure the modulus of the polaritons' momentum after their propagation up to different distances from the excitation spot. To do this, we spatially filter the condensates' PL at different locations along the device and collect the angle-resolved PL of the selected region. The evolution of the polariton momentum as condensates propagate is depicted in Figure 6 (b). The condensates traveling to the central region of the waveguide have initially a momentum of 2.9 $\mu m^{-1}$. This value can be tuned through the power density of the excitation beam. Once they propagate a short distance from the excitation area, the momentum, and therefore the velocity, suffers a significant decay (~25%). A minimum value of 2.2 $\mu m^{-1}$ is obtained at the bend of the waveguide (the bend area is indicated with a gray rectangle). Nonetheless, as the condensate goes over this region, the momentum levels up and remains constant while propagating along the center of the device. The sharpness of the bend acts as an obstacle in the propagation of condensates, slowing down their trajectory. As mentioned above, the initial conditions are easily established, however, the control over the propagation is lost once the bend has been crossed. These results evidence the need of further investigation to improve propagation along more complex polariton circuits.

## 5. Conclusion

We have discussed the polariton dispersion relations in a two dimensional microcavity after its processing to write down waveguides along which polariton condensates are to propagate. We have considered a basic polariton circuit consisting of input and output terminals at the edges of the waveguide, deviating 45º from the longitudinal part of the pathway. The introduction of the 45º turns in the waveguide has a remarkable impact on both the polariton dispersion and propagation. We have observed an asymmetry in the intensity distribution for positive and negative values of $k_x$ and $k_y$ in the linear dispersion relations obtained in the terminals. In the nonlinear regime, when polaritons condense, we have found a slowdown of the polariton condensates' speed as they cross the bend from the input terminal to the coupling region and a multimode emission pattern. The numerical simulations helped to better understand the role played by the symmetry and the edges of the couplers, which produce interference effects between counter-propagating polaritons responsible for the asymmetry in the intensity distributions.

**Supporting Information**

Supporting Information is available from the Wiley Online Library or from the author.

**Acknowledgements**

E. Rozas and J. Beierlein contributed equally to this work. This work has been partly supported by the Spanish MINECO Grant No. MAT2017-83722-R. E.R. acknowledges financial support from the FPI Scholarship No. BES-2015-074708. The Würzburg and Jena group acknowledge financial support within the DFG project PE 523/18-1, KL3124/2-1 and SCHN1376 3-1. The Würzburg group is grateful for support from the state of Bavaria and within the Würzburg-Dresden Cluster of Excellence ct.qmat. A.Y. and I.A.S. thank the Russian Science Foundation for financial support, Project No. 18-72-10110.

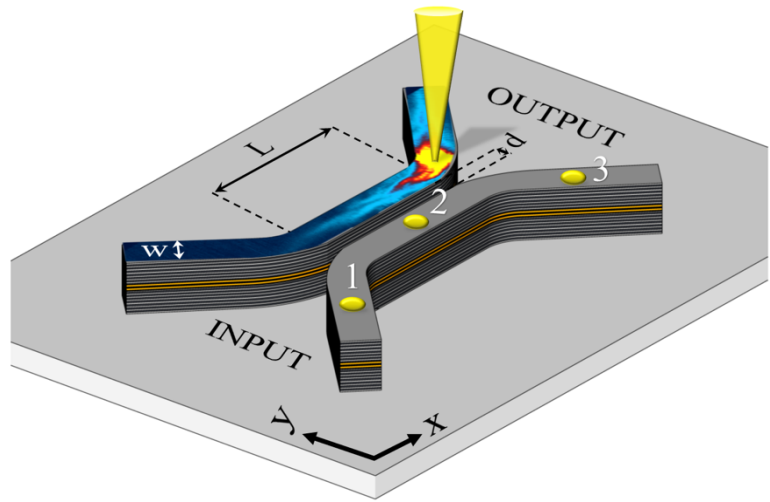

**Figure 1.** Sketch of a coupler device with dimensions defined through the coupling length (L), the separation between the arms (d) and their width (w). Input and output denote the terminals. The different excitation points where the emission has been measured are marked as 1, 2 and 3. As an illustration, the emission, under high excitation conditions, of polariton condensates propagating along the waveguide is depicted when the laser excites the structure in one arm.

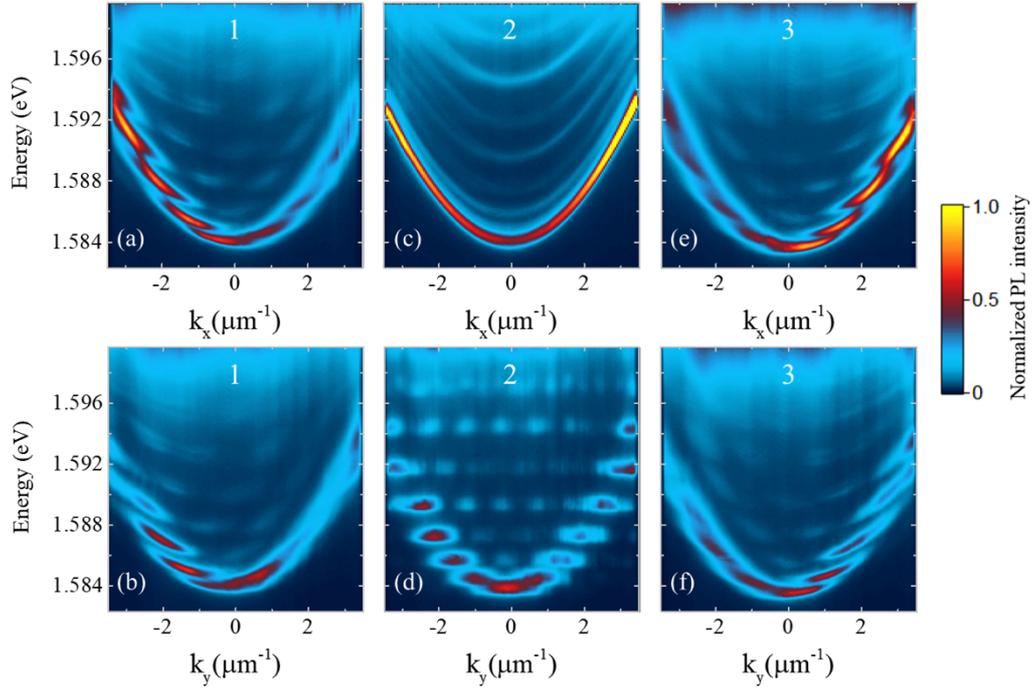

**Figure 2.** Polariton dispersion curves obtained along $k_x$ at locations 1 (a), 2 (c) and 3 (e). The corresponding emissions along $k_y$ are depicted in (b), (d) and (f), respectively. The emissions have been measured at a power density of 0.6 kWcm$^{-2}$.

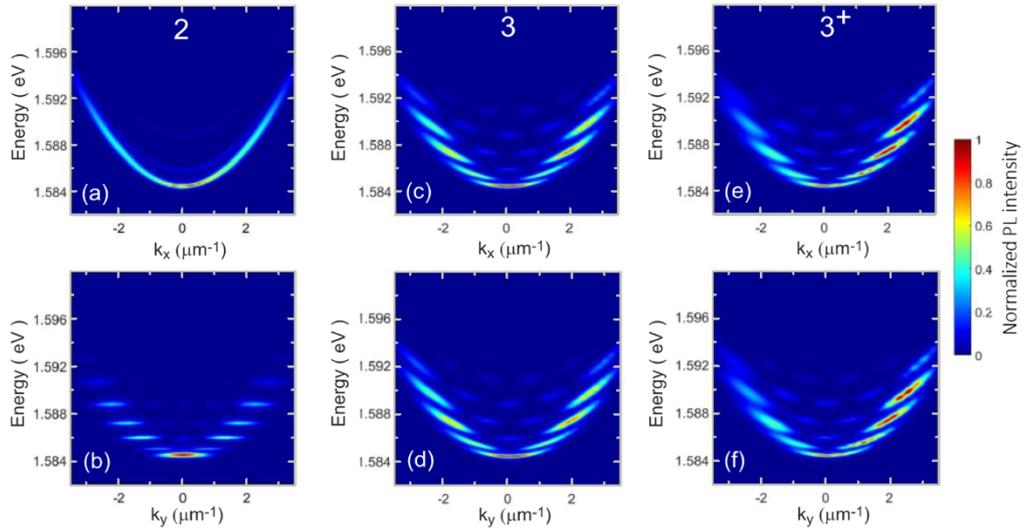

**Figure 3.** Panels (a,b) show numerically calculated dispersion relations for the same excitation spot as for the experiments shown in Figure 2(c, d). The numerically calculated dispersion relations are shown in panels (c),(d) for the waveguide excited at position 3, in the output terminal, corresponding to the experimental dispersions presented in Figure 2(e),(f). Panels (e, f) are obtained for an excitation spot displaced towards the edge of the terminal. The symmetry breaking is clearly seen especially when the waveguide is excited closer to its end (1/3 of the terminal length).

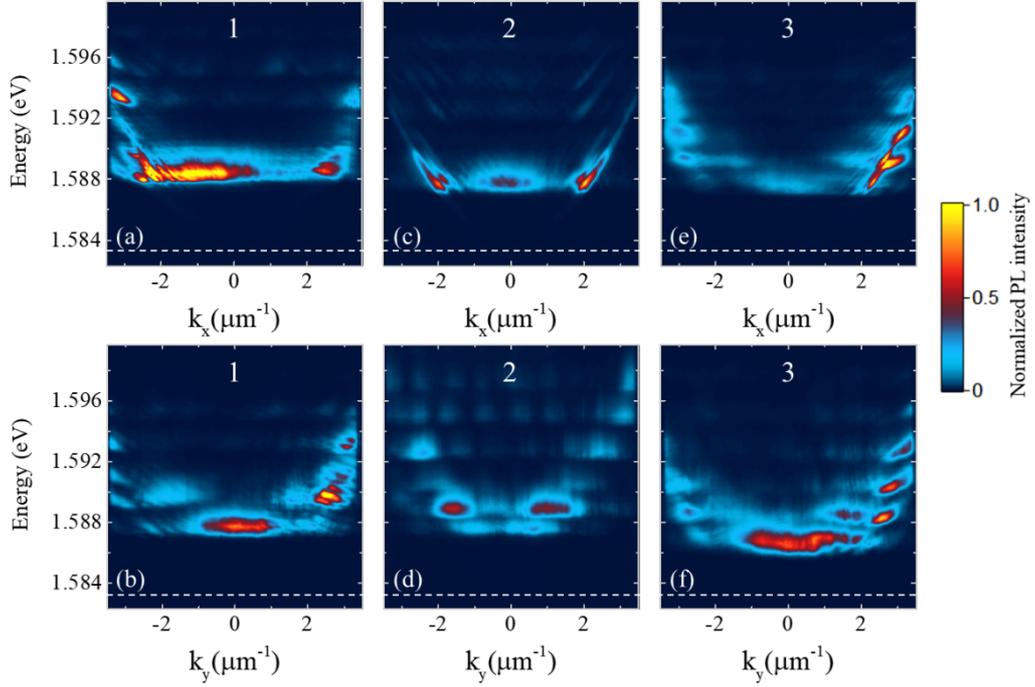

**Figure 4.** Polariton dispersion curves obtained for a higher power density, 24 kWcm$^{-2}$, along $k_x$ at locations 1 (a), 2 (c) and 3 (d). The corresponding emissions along $k_y$ are depicted in (b), (d) and (f), respectively.

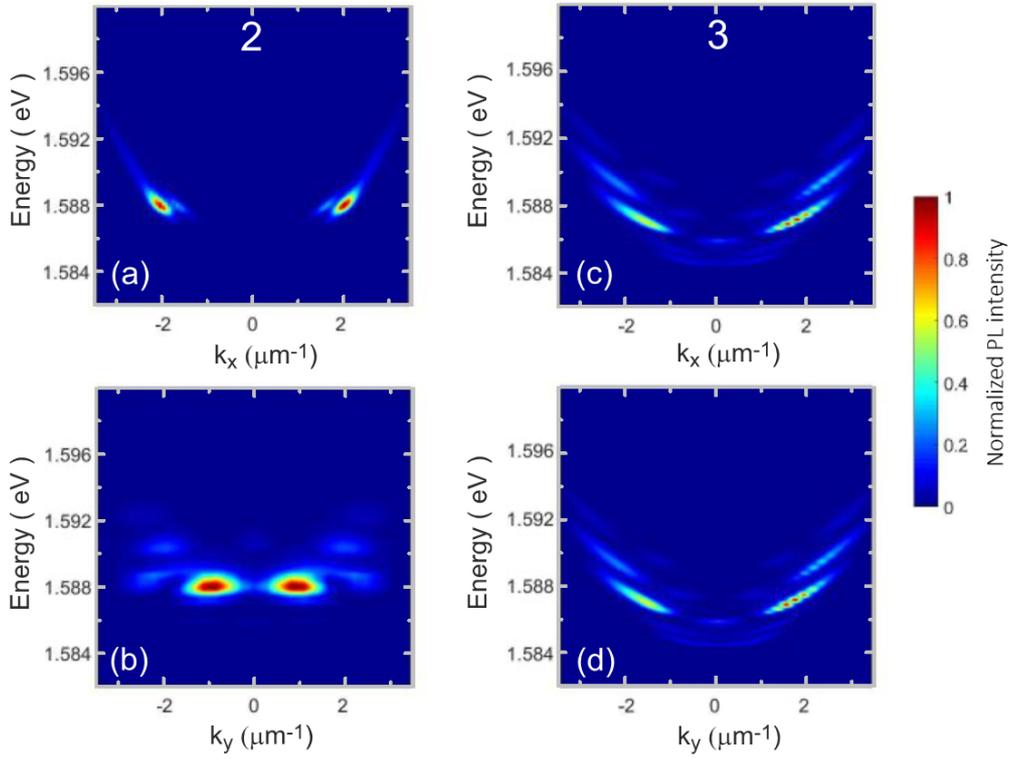

**Figure 5.** Numerically calculated polariton dispersions at above threshold for condensation excitation power density. Panels (a,b) correspond to the experimental dispersions shown in Figure 4 (c,d), while panels (c,d) to those compiled in Figure 4 (e,f).

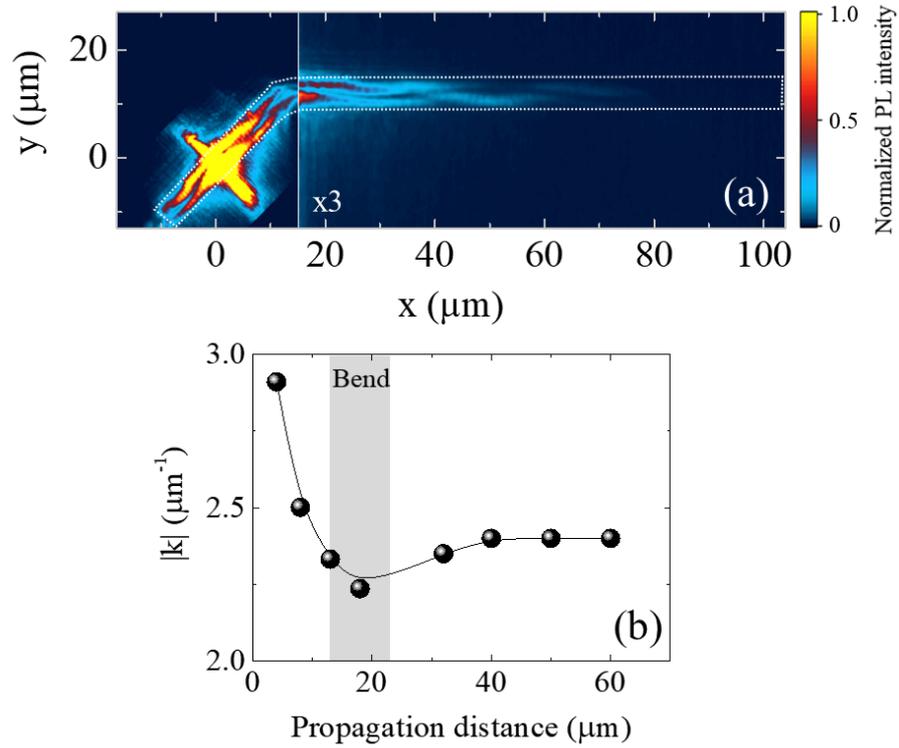

**Figure 6.** (a) Real space map of polaritons traveling along the waveguide. After passing the bend, polaritons follow a zig-zag trajectory. (b) Decay of the polariton condensate's momentum while propagating from the excitation spot at x = 0 μm. A maximum decay is observed at the bend of the waveguide (~ 20 μm).

Table of contents

**Keyword**
polaritons, condensates, microcavities, optical spectroscopy

E. Rozas, J. Beierlein, A. Yulin, M. Klaas, H. Suchomel, O. Egorov, I. Shelykh, U. Peschel, C. Schneider, S. Klembt, S. Höfling, M. D. Martín* and L. Viña

**Impact of the energetic landscape on polariton condensates' propagation along a coupler**

The propagation of polariton condensates along a non-straight circuit is reported. The polariton dispersion relations in different positions along a doubly bent waveguide are measured in both the linear and nonlinear regimes and compared with mean field-based numerical simulations. A slowdown of the polariton speed is found when the particles cross the bend.

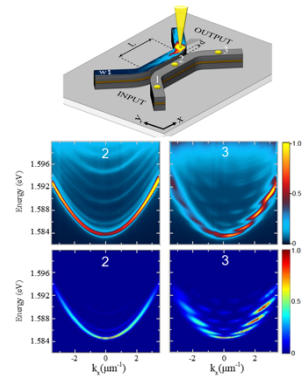

# Supporting Information

**Impact of the energetic landscape on polariton condensates' propagation along a coupler**

Elena Rozas, Johannes Beierlein, Alexey Yulin, Martin Klaas, Holger Suchomel, Oleg Egorov, Ivan Shelykh, Ulf Peschel, Christian Schneider, Sebastian Klembt, Sven Höfling, M. Dolores Martín* and Luis Viña.

It is instructive to see how the emission is distributed in momentum space. In the experiment, the sample is pumped non-resonantly with a pulsed laser beam at 1.664 eV and a power density of 0.6 kWcm$^{-2}$. The emission is filtered at an energy of ~ 1.588 eV.
The excitation spot is situated at the center of the waveguide (position 2) so that the effect of the reflection of the waves from the waveguide edges is negligible. The experimentally measured momentum map is shown in Figure S1 panel (a): the conspicuous bright spots are associated with the excitation of waveguide modes having different transverse structures.
To prove this, and since we are only interested in knowing the energy distribution between the eigenmodes, for the sake of simplicity, we performed numerical simulations exciting the system with a cw source of low intensity and with the same frequency as the detection energy, 1.588 eV. For each of the simulations we took a random distribution of the driving force within the excitation spot. Then the spatial spectra (momentum maps) of the field measured in the light collection spot were calculated and averaged over 100 runs. For each of the runs a new random spatial distribution of the driving force was used. The averaged numerically calculated momentum map is shown in panel (b) of Figure S1.
From a comparison between panels (a) and (b), it is evident that the experimentally measured map and the numerically calculated averaged one are very similar. The brightness of the different spectral patterns vary due to the different efficiency of the excitation of different transverse modes. So, the simple theoretical model reproduces the experimental results well and then using the theory it is easy to check that each of these spectral features corresponds to the excitation of an eigenmode. The spectral patterns with $k_y = 0$ correspond to the fundamental mode, the maxima with other $k_y$ correspond to higher modes having a larger number of field variations along y.

Video S1. Momentum space distribution map of the PL measured at position 2 as a function of the energy. The sample is excited under non-resonant conditions at 1.664 eV and a power density of 0.6 kWcm$^{-2}$. The PL intensity is normalized and depicted in a linear color scale.

Video S2. Momentum space distribution map of the PL measured at position 1 as a function of the energy. The sample is excited under non-resonant conditions at 1.664 eV and a power density of 0.6 kWcm$^{-2}$. The PL intensity is normalized and depicted in a linear color scale.

Video S3. Time-resolved emission of polariton condensates propagating along the output terminal of a coupler. The coupler is pumped in the central region, close to position 2, $(x, y) \simeq (0,0)$ μm. Polaritons are initially ejected from the excitation area and directed towards the bend of the arm, located at (27, 0) μm. They cross the bend at ~18 ps and continue propagating until they reach the edge of the terminal [(72,-32) μm] at ~50 ps. A clear zig-zag trajectory is observed after the condensates pass through the bend, reflecting the impact of introducing a non-straight path on the propagation of the condensates. The emission has been measured by means of a streak camera with a time resolution of 10 ps filtered at an energy of ~1.588 eV, while pumping the structure at 1.664 eV and a power density of 24 kWcm$^{-2}$.

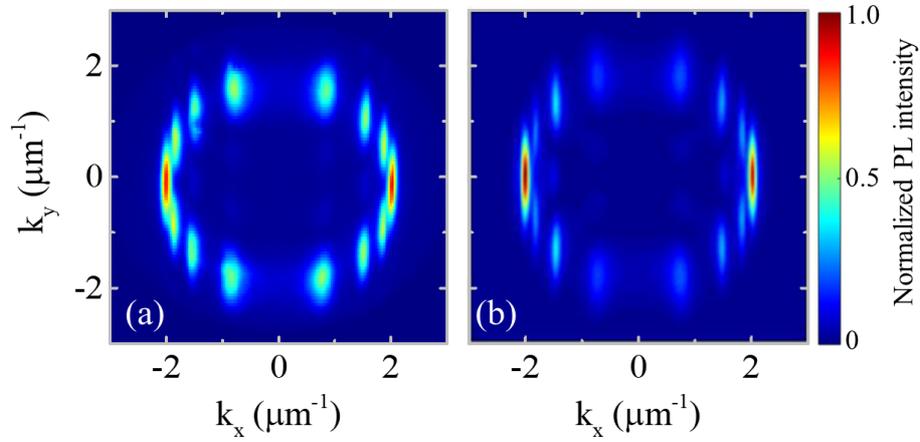

Figure S1. (a) Experimental measurement of the momentum space distribution map at position 2 and filtered at 1.588 eV. The momentum map is measured using a non-resonant pulsed laser beam at 1.664 eV with a power density of 0.6 kWcm$^{-2}$. (b) Corresponding theoretical simulation of the momentum space distribution map for polaritons excited by a cw pump with a frequency corresponding to 1.588 eV; the map is obtained averaging over 100 simulations for different random spatial distributions of the driving force within the excitation spot.